# Optical Control of transverse Motion of Ionization injected Electrons in Laser Plasma Wakefield

Jie Feng,[1,2] Yifei Li,[1] Jinguang Wang,[1] Dazhang Li,[5 *] Changqing Zhu,[1,2] Junhao Tan,[1,2] Xiaotao Geng,[1] Feng Liu,[4] Weimin Wang,[1] Liming Chen,[1,2,3,4 *]

[1]*Beijing National research center of Condensed Matter Physics, Institute of Physics, CAS, Beijing 100190, China*
[2]*University of Chinese Academy of Sciences, Beijing 101408, China*
[3]*Songshan Lake Materials Laboratory, Dongguan, Guangdong 523808, China*
[4]*IFSA Collaborative Innovation Center and School of Physics and Astronomy, Shanghai Jiao Tong University, Shanghai 200240, China*
[5]*Institute of High Energy Physics, CAS, Beijing 100049, China*

## Abstract

We demonstrate an all-optical method for controlling the transverse motion of ionization injected electron beam, by utilizing the transversely asymmetrical wakefield via adjusting the shape of the laser focal spot. When the laser spot shape is changed from the circular to the obliquely elliptic in experiment, the electron beam shape becomes from elliptic to three different shapes. The 3D-PIC simulation results agree well with experiment, and it shows the trajectories of the accelerated electrons change from undulation to helix. Such an all-optical method could be useful for convenient control of the transverse motion of an electron beam which results in synchrotron radiation with orbit angular momentum.

*Corresponding authors:

Liming Chen, E-mail: lmchen@iphy.ac.cn; Dazhang Li, E-mail: lidz@ihep.ac.cn

# I. Introduction

The concept of laser plasma wakefield accelerators (LWFAs) was firstly proposed by Tajima and Dawson [1]. Over the past decades, LWFAs have become increasingly mature and have recently exhibited stability [2], low divergence (mrad) [3], and energy tunability [4] electron bunches with a charge at the pC-level [5]. An electron beam is most efficiently produced in the "bubble" regime [6] which requires the laser pulses that are both intense (normalized vector potential $a_0 > 1$) and short (pulse duration $\tau \leq 2\pi c/\omega_p$, where $\omega_p$ is the plasma frequency). The ponderomotive force of this laser pulses propagating in an underdense plasma pushes the background electrons away from the high intensity regions and drives a relativistic plasma wave. The wave consists of a string of ion cavities (also referred to as bubbles), and the electrons trapped inside can be accelerated by the electrostatic field set up by the separation of electrons and ions. There are many ways of electron capture, including ponderomotive force injection [7], colliding laser pulse injection [8], plasma density gradient injection [9] and transverse self-injection [10, 11]. As for these methods, the injected directions of electrons are hard to control, and these injection processes are not convenient to realize in experiment. In contrast, another method is named ionization induced injection [12-14] which is utilized in this work. Due to the different ionization potential levels of high Z atoms [15, 16] (such as nitrogen), the outer shell electrons will be ionized instantaneously by the rising edge of the laser pulses (98 eV for $N^{+5}$ requires intensity $\sim 2\times 10^{16}$ W/cm$^2$) and pushed away, while the inner shell electrons (552 eV for $N^{+6}$ requires intensity $\sim 1\times 10^{19}$ W/cm$^2$) are ionized close to the peak of the laser intensity envelope. These inner shell electrons hardly exist extra kinetic energy, then they would oscillate in the laser electric-field and slip toward the back of the bubble [17, 18]. The ionization injection is a more controllable method, particularly known for its stability [12, 17, 19]. More importantly, these trapped electrons mainly oscillate along the direction of laser polarization in the ion cavity.

Due to the fact that the plasma wakefield exists a naturally transverse constraint field of tens of GV/m and the radius of plasma bubble is limited to several μm [20], it is difficult to find such a strong external electric-field or magnetic-field for controlling the transverse motion of electron beam in bubble, especially for helical motion. Moreover, Luo et al [21] simulated and acquired the helical motion of electron beam and elliptically polarized radiation by a laser pulses incident with a skew angle relative to the axis of plasma waveguide, but this method was hard to achieve in experiment. Thaury et al [22] utilized one laser pulse to drive an asymmetrical plasma wakefield and the other pulse of colliding injection to realize the helix motion of electron beam, but it is hard for the two laser pulses to overlap in time and space. In addition, He et al [23, 24] also generated the helical motion of electron beam and circularly polarized radiation by using an ultra-intense (PW) circularly polarized laser pulse in near-critical density plasma, but the divergences of electron beam and radiation were very large and the ultra-intense laser facilities (PW) were rare in the world at present [25-28].

In this paper, we proposed a simple all-optical method to control the transverse motion of the ionized injection electron beam by changing the evolution of laser spot shape when propagating and self-focusing in nitrogen gas. We also utilized the three-dimensional particle-in-cell (PIC) simulation to verify our experimental results and analyze the dynamics of electron transverse motion.

## II. Experimental set-up and results

The experiment has been performed at the Key Laboratory for Laser Plasmas of Shanghai Jiao Tong University using the 100 TW laser system, a Ti: Sapphire laser operating at 10 Hz repetition rate with the central wavelength $\lambda_0$ of 795 nm. In the experiment, the system delivered 3 J p-polarized pulses with the duration of 30 fs in FWHM. The experimental setup is shown in Fig.1. The laser beam was focused by an $f/20$ off-axis parabola (OAP) mirror to a vacuum spot size closing to Guassian intensity distribution with FWHM diameter of 30 μm containing 30% energy. And by adjusting the OAP posture to the optimal position, the intensity distributions and spot shapes at, before and after the focal plane are all approximately circular, as shown in Fig. 1(a). So the vacuum-focused laser intensity can reach up to $6.5 \times 10^{18}$ W/cm$^2$, for which the corresponding normalized vector potential $a_0$ is ~1.7. The plasma target was formed using a 1.2 mm×4 mm supersonic gas jet, which can generate well-defined uniform gas density profiles in the range of $1\times10^{17}$ cm$^{-3}$ to $3\times10^{19}$ cm$^{-3}$ by changing the gas stagnation pressure [29-31]. And the laser focal plane was located at the front edge of the gas jet. The experimental results were obtained using pure nitrogen for ionization induced injection, and the gas stagnation pressure was set at 1.2 bar corresponding to the background plasma density of $6\times10^{18}$ cm$^{-3}$, considering the outer shell electrons of nitrogen were fully ionized. A top-view [32] system consisting of a 14-bit charge-coupled device (CCD) and a low-pass filter, was used to monitor the interaction position and the length of time-integrated plasma channel, as result shown in Fig. 1(c). The electron beams emitted from the plasma channel were detected by a fluorescent DRZ screen coupled with a 16-bit EM-CCD. The laser pulses were blocked by a beryllium window with the thickness of 350 μm. However, in order to control the transverse motion of electron beam by using asymmetric transverse wakefield, we adjusted the OAP postures of sway, pitch and rotate to generate the laser intensity distribution of oblique 45° elliptic and oblique 135° elliptic for the before and after focal spot respectively, as shown in Fig. 1(b).

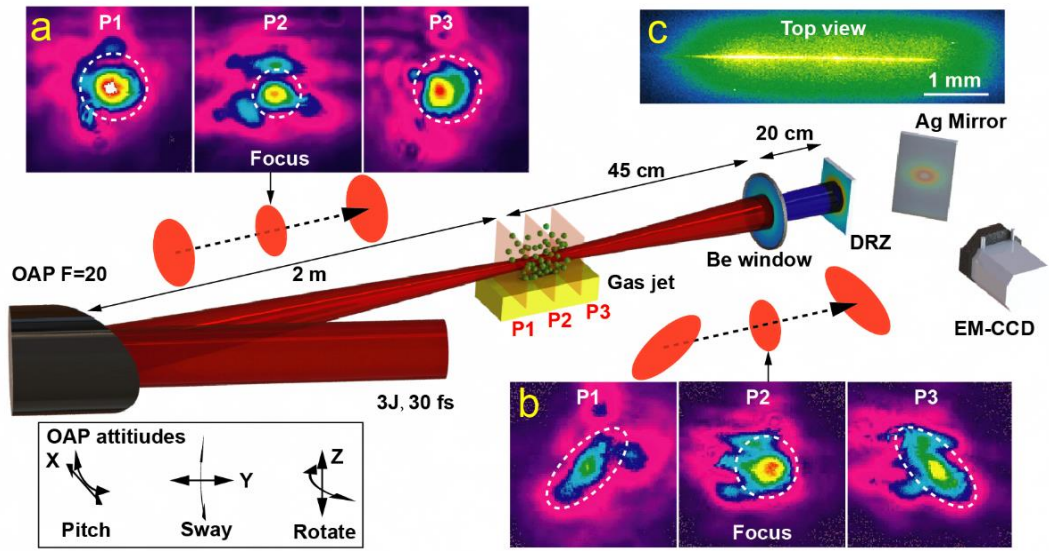

**Figure 1.** Experimental setup. (a) measured laser intensity distribution before, at and after the focal spot in the case of perfect focus situation respectively, and (b) measured laser intensity distribution after adjusting the attitudes of OAP, and (c) is the top-view image of the plasma channel.

The experimental results of electron beam spots driven by laser pulses with the two kinds of focal spot shape are shown in Fig. 2. When the posture of OAP is in the optimal position, the typical shape of these electron beam spots is ellipse with long axis along the horizontal direction (y-direction), as shown in the fourth column (0°). As for the case of Fig. 1(b), the experimental results are shown in Fig. 2, which demonstrate three different types of ellipse with long axis along the direction of 45° 90° and 135° respectively. To explain the results and analyze the dynamics of electron beam in plasma wakefield, we also carried the 3D-PIC simulations in the next sections.

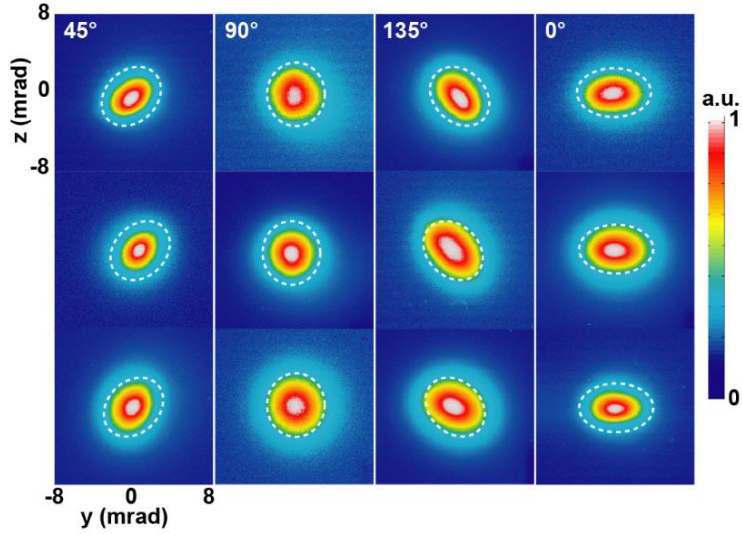

**Figure 2.** Electron beam spatial distribution.

## III. PIC simulation and results

The 3D-PIC simulations were carried out using the KLAPS code [33, 34], and the tunnel-ionization model was adopted for field ionization. The simulation box size was 50 ×60 ×60 μm$^3$ with 1000 ×600 ×600 cells at x, y and z directions respectively, and one cell contains four macro particles. Moreover, a third order time interpolation for the magnetic field was used in the simulation. A p-polarized (y-direction) 800 nm laser pulses with $a_0$ = 1.7 was focused to a radius of 15 μm at x= 50 μm after the front edge of the nitrogen gas. The pulse had a Guassian transverse profile and sin-square longitudinal shape with pulse duration FWHM of 30 fs. The neutral nitrogen longitudinal profile had a 100 μm up-ramp followed by a 2 mm long plateau with uniform density of 6×10$^{17}$ cm$^{-3}$, corresponding to the background plasma electron density of 6×10$^{18}$ cm$^{-3}$.

If a laser pulses with focus situation shown in Fig. 1(b) is propagating and self-focusing in the plasma, the shape of the laser spot will change from an oblique 45° ellipse, to a circular and then to an oblique 135° ellipse. This process will be repeated until the laser pulses cannot be self-focusing in the plasma. Therefore, in order to study the process of this laser pulses propagating and self-focusing in the plasma, and the influence of the asymmetrical laser spot on the plasma wakefield acceleration, the asymmetrical spot intensity distribution before the focal plane was set according to the measured intensity distribution in the experiments (as show in Fig. 1(b)), and the electric-field distribution is expressed as:

$$E(x,y,z) = E_0 \cdot \sqrt{w_0/rs(x)} \cdot \exp\left(-\left(\frac{(y \cdot \cos\theta - z \cdot \sin\theta)^2}{2} + \frac{(y \cdot \cos\theta + z \cdot \sin\theta)^2}{0.5}\right)\bigg/rs(x)^2\right) \quad (1)$$

Where $rs(x) = w_0 \cdot \sqrt{1-(x-x_0)^2/z_R^2}$, $x_0$ is the longitudinal position of focal plane, and $z_R$ is the Rayleigh length. $\theta = 45°$ is the rotation angle (clockwise direction) of the long axis of ellipse shape.

The simulation results are shown in Fig. 3. For the case of symmetrical laser spot, the electron beam in the plasma bubble demonstrate that it has a bigger divergence at the y-direction (laser polarization direction) than the z-direction because of ionization injection, as shown in Figs. 3(a, b). Moreover, the simulation results also demonstrate the cross-sections (YZ plane) of bubble, and the electron beam mainly oscillate in the plane of XY, as shown in Figs. 3(e, f). However, for the case of asymmetrical laser spot, the divergence of electron beam at the y-direction is reduced, but increasing obviously at the z direction as shown in Figs. 3(c, d). Moreover, the shape evolution of the cross-section is similar to the laser spot, due to the ponderomotive force $F_p \simeq -m_e c^2 \nabla(a^2/2)$ expels electrons from the intense region of the laser pulses and excites plasma bubble. In addition, the electron beam transverse motion is different from the case of symmetrical laser spot, as shown in Figs. 3(h-j).

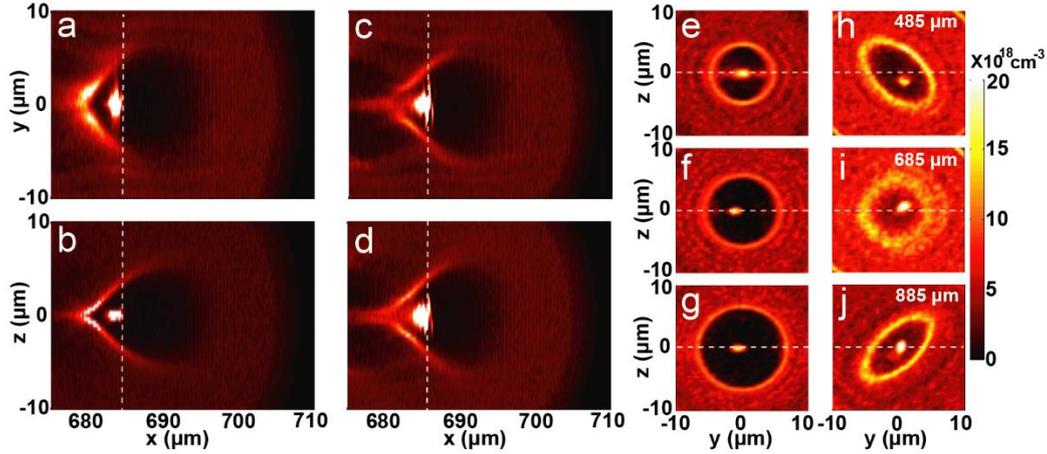

**Figure 3.** Laser plasma wakefield acceleration in PIC simulations. (a, b) is the slice of plasma bubble in the plane of XY and XZ respectively, driven by the symmetric laser focal spot, and (c, d) are the slices driven by the asymmetric laser focal spot, and (e-g) are the cross-sections (YZ plane) of plasma bubble at different propagation positions, corresponding to the case of symmetric focal spot, and (h-j) corresponding to the case of asymmetric focal spot at different propagation positions.

The simulation electron spots are shown in Figs. 4(a-d), and Figs. 4(a), 4(b) and 4(c) are electron spots at different propagation distance of laser pulses, corresponding to Figs. 3(j), 3(i), and 3(h) respectively. The electron spot shape is also similar to the cross-section shape of plasma bubble, and the three typical simulation spot shapes agree well with the experiment results. Moreover, according to the simulation, the laser propagation distance for the electron spot changing from one shape to the another is about 200 μm, as shown in Figs. 3(h-j). Therefore, the experiment results were acquired at the same condition, the electron spot occurs three different shape, which is mainly caused by the different electron injection position and propagation distance. However, for the case of symmetrical focal spot, the shapes of the electron beam of experiment and simulation are always ellipse with long axis along the laser polarization direction, as shown in Fig. 3 and Fig. 4 (d).

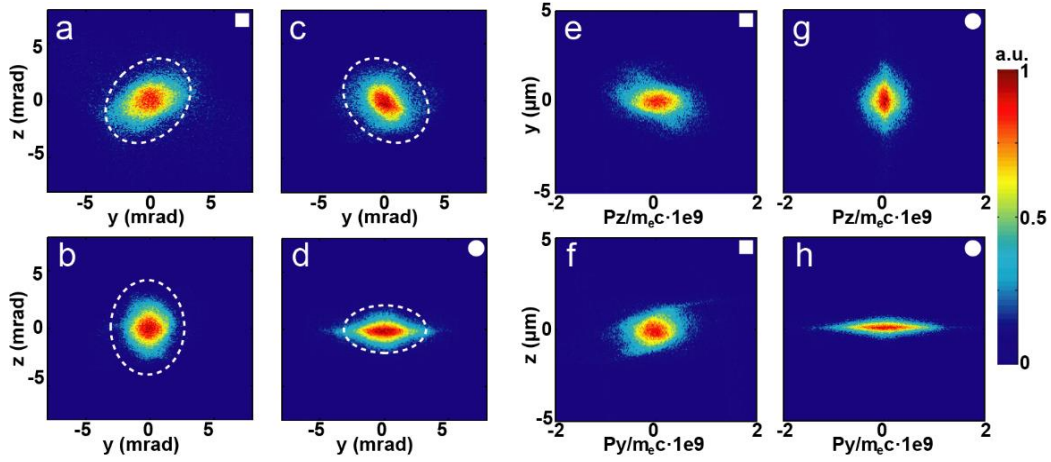

**Figure 4.** Electron beam spots of PIC simulation. (a-c) electron beam spots driven by a 45° oblique elliptical laser spot at different laser propagation distances in nitrogen, and (d) corresponds to a circular laser spot, and (e, f) is the phase space of $P_y$-z and $P_z$-y distribution respectively, corresponding to the electrons of Fig. 4(a), and (g, h) correspond to the Fig. 4(d).

## IV. Discussions

To discuss the influence of the asymmetrical plasma wakefield on the transverse motion of accelerated electron beam in plasma bubble, the phase spaces of $P_y$-z and $P_z$-y corresponding to the electrons of Figs. 4(a) and 4(d) are also analyzed as shown in Figs. 4(e, f) and Figs. 4(g, h) respectively. Electrons in symmetrical plasma wakefield has larger momentum at the y-direction than the z-direction, as shown in Figs. 4(g, h), resulting the shape of the electron beam tends to be an ellipse, as shown in Figs. 4(d). However, for the asymmetrical plasma wakefield, the maximum momentum at y-direction is approximately equal to the z-direction, as shown in Figs. 4(e, f). Therefore, a majority of electrons have experienced a strong force at the z-direction during the evolution of plasma bubble.

In addition, the transverse forces felt by the electrons on a cross section are also analyzed. Figs. 5(a, d) are the cross section of the plasma bubble corresponding to the cases of Figs. 4(c) and 4(d) respectively, and Figs. 5(b, e) and Figs. 5(c, f) are the corresponding force distribution of $f_y/|q|$ and $f_z/|q|$ expressed as:

$$f_y(y,z)/|q| = -E_y(y,z) - v_x \cdot B_z(y,z) \qquad (2)$$

$$f_z(y,z)/|q| = -E_z(y,z) + v_x \cdot B_y(y,z) \qquad (3)$$

Where $E_y$ and $B_z$ are the self-generated fields in plasma bubble, and $v_x$ is the speed of electron along the direction of laser propagation. For the symmetrical laser spot, the transverse plasma wakefield is also geometrically symmetrical, as shown in Figs .5(e, f). The force at the z-direction felt by the electron beam is almost equal to zero in the polarized plane of the maximum laser intensity, as shown in Fig .5(f). Generally, ionized injection electrons would have residual momentum along the electric-field direction of laser pulses. Thus, the injected electrons mainly oscillate along the y-direction driven by the force of $f_y$, and the transverse field gradient could reach 100 GV/m, as shown in Fig. 5(e). While, the oblique elliptic laser spot will destroy the symmetries of the transverse wakefield, as shown in Figs .5(b, c). Although the ionized injection electron beam initially oscillates in the plane of laser polarization, it would gradually deviate from the direction as shown in Fig. 5(a). It is because that the electron beam experienced a strong force of $f_z$ and the field gradient could be up to 50 GV/m. Moreover, the direction of resultant force tends to along the long axis of the cross section of bubble, resulting in a similar electron beam shape as the cross section of bubble. And, the direction of resultant force would change with the evolution of the cross-section of wakefield. Thus, these electrons would rotate around the propagation direction and form a spiral electron beam.

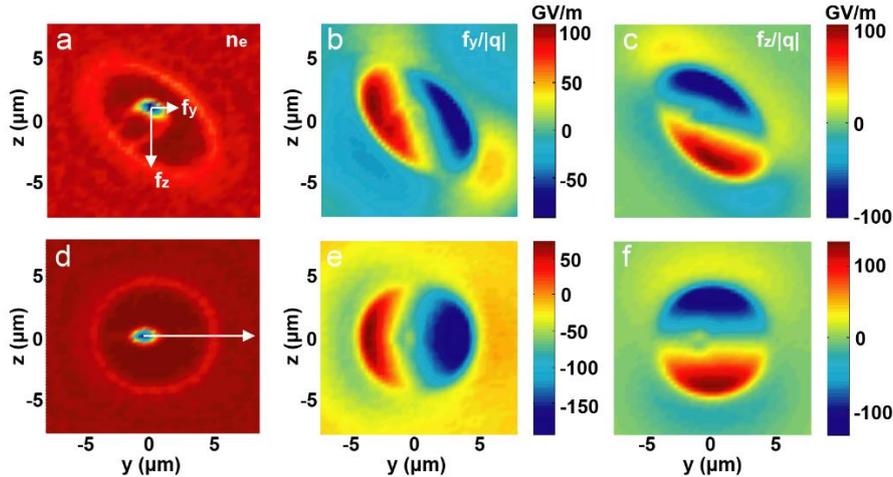

**Figure 5.** Analysis of transverse force for the electrons in plasma wakefield. (a, d) are the cross sections of plasma bubble corresponding to elliptical spot and circular spot respectively. (b, c) is the transverse force at the direction of y and z respectively of Fig. (a); (e, f) is the same of Fig. (d).

To compare the motion of electron beam in the acceleration process for the two cases, we tracked 20 electrons respectively, and their trajectories are shown in Figs. 6(a) and 6(b). For the case of symmetrical laser spot, these electrons mainly oscillate in the plane of XY with the maximum amplitude of ~5 μm, and the amplitude in the plane of XZ is about 0.5 μm. However, For the case of asymmetrical laser spot, these electrons present helical motion, and their oscillation amplitudes are both about 3 μm in the plane of XY and XZ. This kind of helical motion is propitious to generate circularly or elliptically polarized synchrotron radiation.

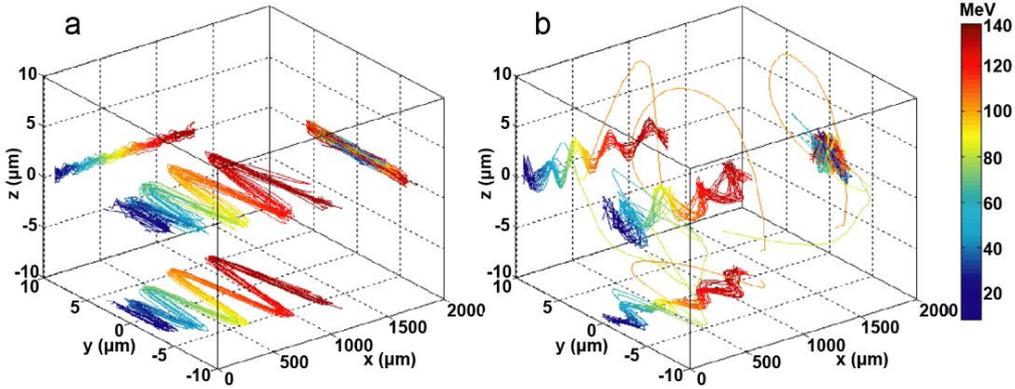

**Figure 6.** Electrons' trajectories driven by circular laser spot and elliptical laser spot respectively.

## V. Conclusion

In conclusion, a simple method of controlling the transverse motion of an electron beam in plasma bubble is presented. Laser pulses with a power of 100 TW drive plasma wakefield and accelerate an electron beam in the regime of ionization injection. The transverse motion of the accelerated electron beam can be controlled by changing the intensity distribution of laser focal spot via adjusting the posture of OAP. When the shape of laser focal spot changing from a circular to an obliquely elliptic, the geometric symmetry of the transverse force in plasma bubble is changed, resulting in the motion of electron beam changing from the structure of undulation to helix. Moreover, the profile of electron beam also changes with the laser focal spot's shape. The experimental results are verified by the 3D PIC simulations. Such a method is expected to conveniently control the transverse motion of electron beam in the wakefield and generate circularly polarized synchrotron radiation.

## Acknowledgments

This work was supported by the National Key R&D Program of China (2017YFA0403301), the National Natural Science Foundation of China (11721404, 11805266), the Chinese Postdoctoral Science Foundation (Y9BK014L51), the Key Program of CAS (XDB17030500), and the Science Challenge Project (TZ2018005).